\documentclass[11pt]{article}
\usepackage{moriond,epsfig}

\def\Journal#1#2#3#4{{#1} {\bf #2}, #3 (#4)}

\def\PLB{{\em Phys. Lett.}  B}

\def\ZPC{{\em Z. Phys.} C}
\newcommand{\SKI}{\mbox{\ttfamily SK-I}}
\newcommand{\JETSET}{\mbox{\ttfamily JETSET}}
\newcommand{\DELPHI}{\mbox{\ttfamily DELPHI}}
\newcommand{\LEP}{\mbox{\ttfamily LEP}}
\newcommand{\LEPtwo}{\mbox{\ttfamily LEP2}}
\newcommand{\SLD}{\mbox{\ttfamily SLD}}

\def\be{\begin{equation}}
\def\ee{\end{equation}}
\def\bea{\begin{eqnarray}}
\def\eea{\end{eqnarray}}

\newcommand{\qqb}{{q{\bar q'}}}
\newcommand{\qbq}{{{\bar q}q'}}
\newcommand{\qqqq}{{\mathrm{ \qqb \qbq}}} 
\newcommand{\WW}{{\mathrm{ W^{+} W^{-} }}}

\begin{document}
\vspace*{4cm}
\title{Possible reduction of the total uncertainty on the $m_W$ measurement
at \LEPtwo\ \footnote{Talk prepared for Rencontres de Moriond XXXVII, electroweak session (Forum for Young Researchers).}}

\author{ J. D'HONDT \footnote{Work supported by IWT-Vlaanderen.} }

\address{Vrije Universiteit Brussel, IIHE, Pleinlaan 2,1050 Brussel, Belgium}



\maketitle\abstracts{An alternative W mass estimator in 
$e^+e^- \rightarrow \WW \rightarrow \qqqq$ events at \LEPtwo\ is designed
to optimize the balance between the statistical uncertainty and the 
systematic uncertainty due to a possible Colour Reconnection effect. 
The preliminary result for the total uncertainty on the W mass in this 
channel is roughly 30\% lower then those obtained with the standard 
estimators, based on the \SKI\ implementation of Colour Reconnection. 
Also an indirect measurement of the \SKI\ Colour Reconnection 
model parameter $\kappa$ is inferred from the difference between both W mass
estimators.}

\section{Status}

A measurement of the mass of the W boson has been performed by all \LEP\
Collaborations \cite{masspaper}, this by the method of direct reconstruction 
of the process $e^+e^- \rightarrow \WW \rightarrow 
\qqqq$, the so-called four jet channel.
The uncertainty on the \LEP\ combined $m_W(4q)$ value however is dominated by
systematical rather then statistical uncertainties. The major component
comes from the assumption in the reconstruction of the event that particles 
from the decay of different W bosons are independent in the 
non-pertubative fragmentation phase of the process. Physics phenomena like 
Colour Reconnection (CR), gluon exchange between fragmentation products, 
could break this assumption. Neither its theoretical 
knowledge nor its existance is well established in these processes.

\section{Proposal}

From the \SKI\ phenomenological model \cite{sk1} which is implemented in the \JETSET\ 
fragmentation scheme, we observe that mostly low momentum particles and 
particles in inter-jet regions are affected by CR and hence influencing the 
measurant, $m_W$. Therefore we could design an alternative analysis which is
neglecting those particles in the reconstruction of the momenta of the four
primary partons, hence decreasing the systematical uncertainty on $m_W$ due
to CR but meanwhile increasing the statistical uncertainty because we neglect
part of the information content of the event. 

Within the standard \DELPHI\ $m_W$ estimator \cite{delphi} an iterative procedure
was used within each pre-defined jet to find a stable direction of a cone 
excluding some particles in the calculation of 
the jet momentum. Starting with the direction of the original jet, the jet 
direction was recalculated only from those particles which have a opening 
angle smaller then $R_{cone}$ with this original jet. This process was 
iterated by constructing a second cone (of the same opening angle) around 
this new jet direction and the jet direction was recalculated again. 
The iteration was continued until a stable jet direction was found. The 
obtained jet momenta were rescaled to conserve the invariant mass of the jets.

\begin{figure}
\begin{center}
\epsfig{figure=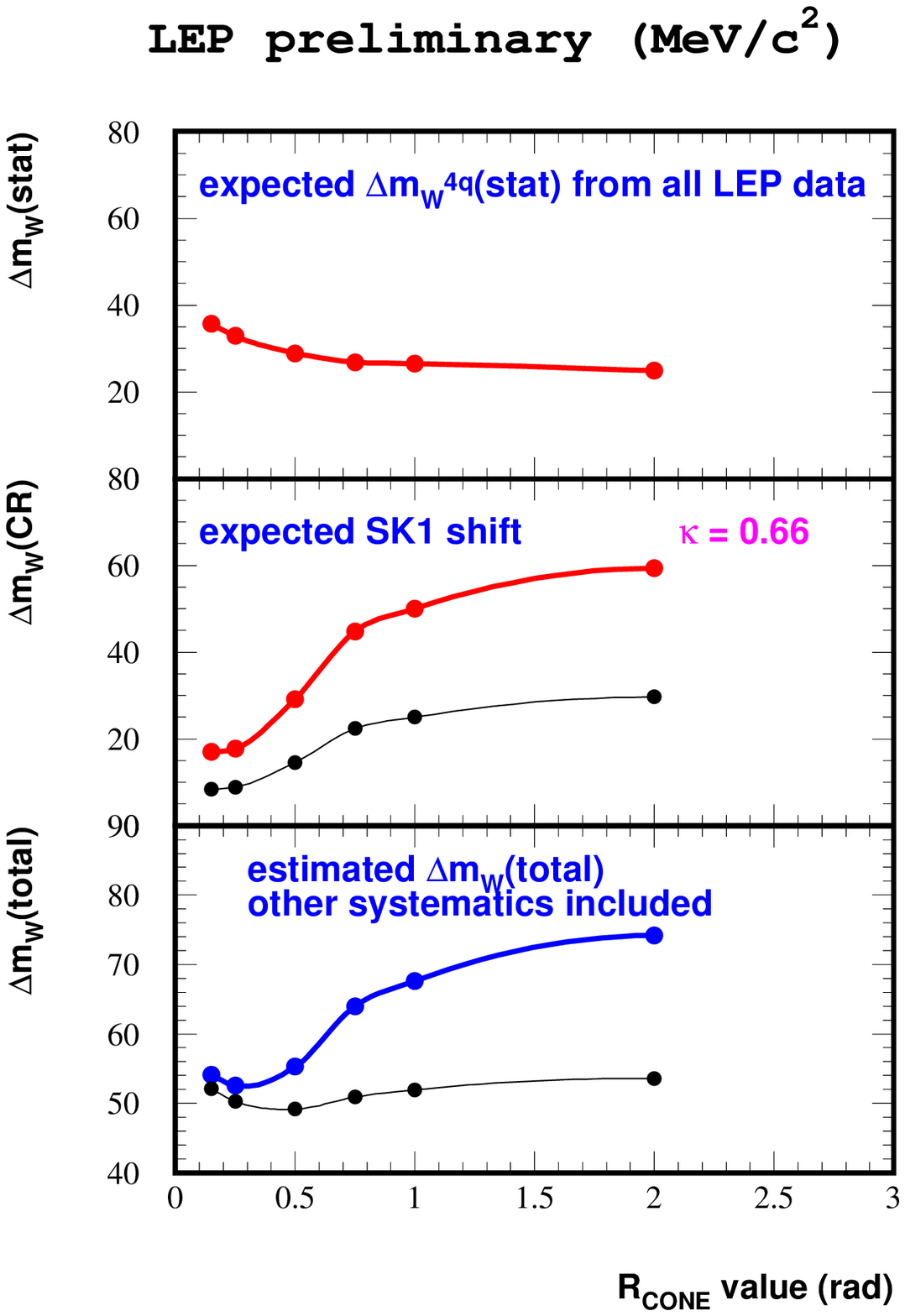,height=9cm}
\epsfig{figure=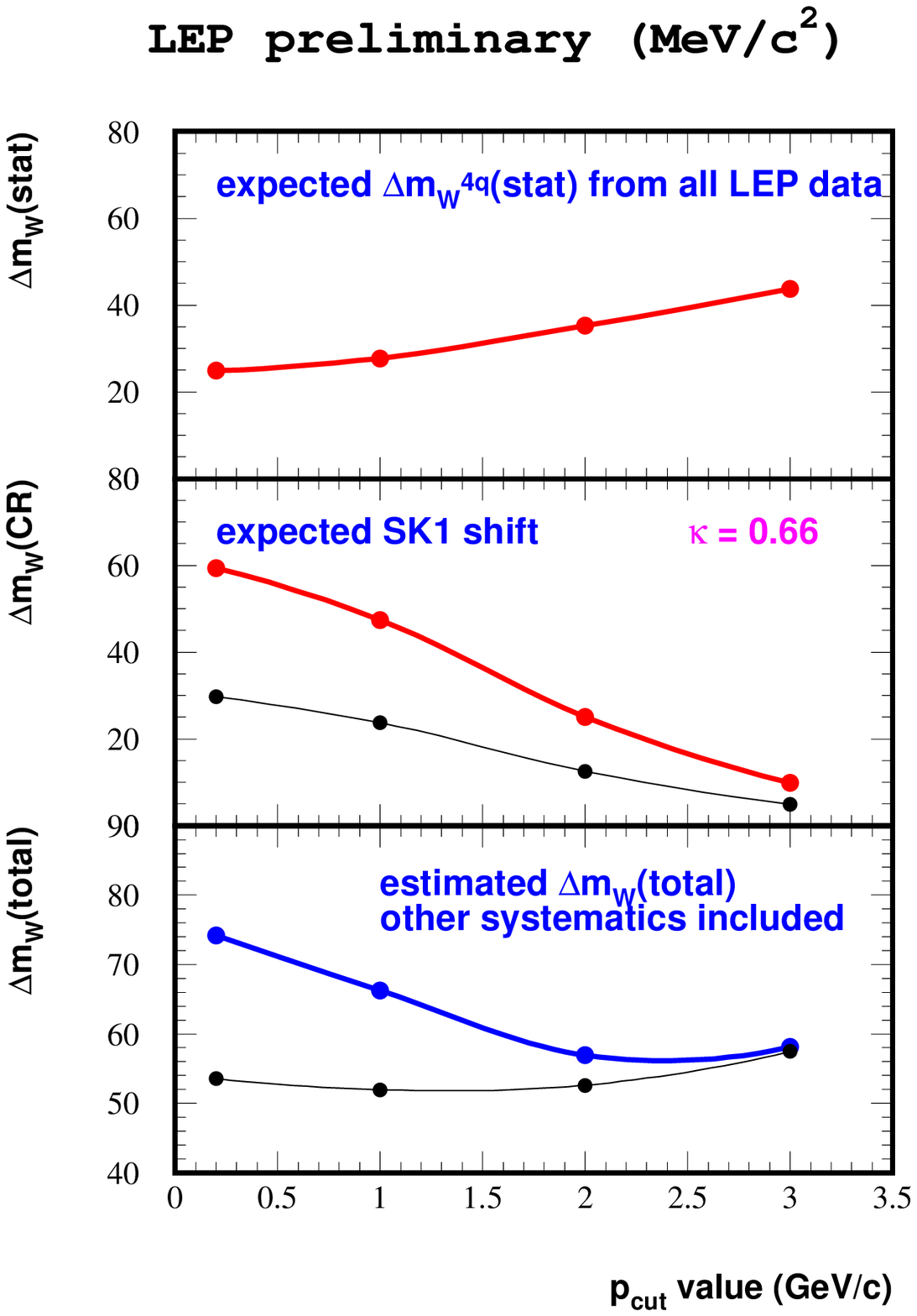,height=9cm}
\end{center}
\vspace{-0.7cm}
\caption{The left figure shows the dependency of the total uncertainty on $m_W$
and its major components as function of the opening angle of the cone, assuming
that all other systematics are invariant under change of $R_{cone}$. The black
line indicates the uncertainty if half of the shift is applied and the other 
half is quoted as systematic uncertainty. The right figure shows the same as function 
of the $p_{cut}$ value.}
\label{fig:mw}
\end{figure}

The results from Monte Carlo studies using the \SKI\ Colour Reconnection 
model with fixed parameter $\kappa=0.66$ predict a decrease of 
the total \LEPtwo\ uncertainty on $m_W(4q)$ when using the appropriate alternative 
analysis ($R_{cone} \simeq 0.5 rad$) of around 30\%, see Figure~\ref{fig:mw}.
Those results are compared with a $m_W$ estimator where simply all particles
with momentum below $p_{cut}$ are rejected when calculating the jet kinematics.

An indirect measurement of \SKI\ model parameter $\kappa$ 
is possible from the direct measurement of the difference in 
reconstructed $m_W$ between the CR-sensitive standard and CR-not-sensitive 
alternative analysis. From Monte Carlo studies we observe that the statistical
significance of this measurement is comparable with the one 
obtained from the direct measurement. Systematics can however become important.

\section{Summary}

Alternative exclusive $m_W$ estimators can be designed which have a
30\% smaller total uncertainty then the standard ones used by the \LEP\
Collaborations. Also information can be inferred about parametrized Colour 
Reconnection models by studying the difference between the standard and the
proposed alternative $m_W$ estimators.

\end{document}